\documentclass[pra,twocolumn]{revtex4}
\usepackage[all]{xy}
\usepackage{amssymb, amsmath}
\usepackage{accents}

\usepackage{graphicx}
\usepackage[caption=false]{subfig}
\begin{document}

\title{Exact ground-state correlation functions of an  atomic-molecular \\ boson conversion model }
\author{Jon Links and Yibing Shen\\ }
\affiliation{School of Mathematics and Physics, The University of Queensland, Brisbane, QLD 4072, Australia}
\begin{abstract}
We study the ground-state properties of an atomic-molecular boson conversion model through an exact Bethe Ansatz solution. For a certain range of parameter choices, we prove that the ground-state Bethe roots lie on the positive 
real-axis. We then use a continuum limit approach to obtain a singular integral equation characterising the distribution of these Bethe roots. Solving this equation leads to an analytic expression for the ground-state energy. The form of the expression is consistent with the existence of a line of quantum phase transitions, which has been identified in earlier studies. This line demarcates a molecular phase from a mixed phase.  Certain correlation functions,  which characterise these phases, are then obtained through the Hellmann-Feynman theorem.     
\end{abstract}

\maketitle

\section{Introduction}

Models of interacting bosons provide wide scope for investigations into quantum properties of many-body systems. Developments in the high-precision control of systems at ultracold temperatures has allowed for clinical comparison between the results of theory and experiment. A classic early example is the Newton's cradle experiment \cite{kww06}, illustrating the effect of dimensionality on the dynamics of an ultracold system. The observations of this work were later successfully explained in terms of integrability \cite{rdyo07}. 

In contrast to fermionic systems, models for bosonic systems may be formulated with very few degrees of freedom, while still accommodating large particle numbers. This property allows for many models to be constructed which are readily amenable to in-depth investigations of many-body properties under a range of techniques, including semi-classical and mean-field analyses \cite{hmm03,ab06,hlxl13,ly15,zzw15,gkr16},  variational methods \cite{lf11}, algebraic approaches \cite{bf08,sbhmf15}, and numerical studies \cite{cf12,dbskb15} to name a few.

The present study concerns a boson conversion model involving only two bosonic modes, one associated with an atomic degree of freedom and another with a homonuclear molecular degree of freedom. This model, including some specialised cases, has been the subject of several earlier studies \cite{vya01,kgv02,zlm03,zlgm03,stfl06,lymfl09,sflmd10,kv11,cwy12,sxy13,sl15,ggr15}. In particular, a quantum phase transition boundary line was identified in \cite{stfl06} through an analysis of the semi-classical equations of motion. It was later shown in \cite{sflmd10} that signatures of the quantum phase transition, in the repulsive case, were evident in calculations of quantities such as entanglement and fidelity.    
Below it will be seen that this quantum phase transition line is confirmed in an analysis of an exact Bethe Ansatz solution of the model. First, an explicit expression for the ground-state energy will be derived. An important property of the model is that the exact solution is valid without restriction on the coupling parameters. This then  facilitates the use of the Hellmann-Feynman theorem to compute a range of expectation values in an exact, analytic manner. Illustrative examples will be provided through calculation of the expectation value of the atomic fraction, and the fluctuations of this quantity. 
 
The model and the exact solution are presented in Sect. II. Sect. III establishes the location of the roots of the Bethe Ansatz equations associated to the ground state in the repulsive case. Specifically, these roots all lie on the positive real-axis of the complex plane, and there can only be one set of solutions with this property. In Sect. IV a continuum limit approximation is applied to the Bethe Ansatz equations, which is appropriate to study the system in the limit of large particle number. Through this approach equations are derived for the Bethe root density and ground-state energy. An asymptotic analysis in this large particle number limit is undertaken in Sect. V, which confirms the existence of a  line of quantum phase transitions which have been identified in earlier studies. The main results of the study are presented in Sect. VI, where expressions for ground-state correlation functions are presented. These are exact in the limit of infinite particle number. Conclusions are presented in Sect. VII.

\section{The Hamiltonian and exact solution}

The model for ultracold, interacting, atomic and molecular bosons is given by the Hamiltonian  \cite{zlgm03,stfl06,lymfl09,sflmd10,lf11,cwy12}
\begin{align}
H=U_{aa}N^2_a&+U_{bb}N^2_b+U_{ab}N_a N_b+\mu_a N_a +\mu_b N_b \nonumber \\
\qquad\qquad  &+\Omega\left(a^\dagger a^\dagger b + b^\dagger a  a  \right)
\label{ham}
\end{align}
where $a,\,a^\dagger$ are the annihilation and creation operators for an atomic mode, and $b,\,b^\dagger$ are the annihilation and creation operators for a diatomic, homonuclear molecular mode. These satisfy the canonical commutation relations
$$\left[ a,\,a^\dagger\right]=\left[b,\,b^\dagger\right]=I$$
where $I$ denotes the identity operator. Moreover
$$\left[ a,^{\phantom{\dagger}}b\right]=\left[a,\,b^\dagger\right]=\left[ a^\dagger,b\right]=\left[a^\dagger,\,b^\dagger\right]=0.$$
The parameters $\mu_i$
are chemical potentials for species $i$ and $\Omega$ is the amplitude for the interconversion of atoms and molecules. The parameters $U_j$ are scattering couplings, taking into account atom-atom ($U_{aa}$), atom-molecule ($U_{ab}$), and molecule-molecule ($U_{bb}$) interactions. 
The Hamiltonian commutes with the total atom number
$N=N_a+2N_b$ where $N_a=a^\dagger a$ and $N_b=b^\dagger b$. 

The limiting case $U_{aa}=U_{ab}=U_{bb}=0$ is the model introduced in \cite{vya01}, for which a Bethe Ansatz solution was given in \cite{zlm03}. This solution was extended for the general Hamiltonian (\ref{ham}), through use of the Quantum Inverses Scattering Method and algebraic Bethe Ansatz, in \cite{zlgm03}. More recently an alternative solution was provided in \cite{sl15} using a differential operator correspondence. It is this latter form of solution which will be utilised in the analysis below.
Set
\begin{align*}
A&=4U_{aa}-2U_{ab}+U_{bb} , \\
B&= 4(k+1) U_{aa}+(2M-k-2)U_{ab}+(1-2M)U_{bb} \\
&\qquad +2\mu_a-\mu_b , \\
C&= k^2U_{aa}+kMU_{ab}+M^2U_{bb}+k\mu_a+M\mu_b .
\end{align*}
%
The energy eigenvalues are given by 
\begin{align}
E&=AM(M-1)+BM+C-\Omega\sum_{j=1}^M v_j 
\label{nrg}
\end{align}
where 
\begin{align}
\frac{Bv_j+\Omega(4k+2-v_j^2) }{Av_j^2 +4\Omega v_j }
&=\sum_{k\neq j}^M\frac{2}{v_k-v_j}
\label{bae}
\end{align}
and $M = (N - k)/2$ where $k = 1$ for odd $N$ and $k = 0$ for even $N$.

\section{Ground-state roots of the Bethe Ansatz equations}

In \cite{ggr15} an approach was taken to express the mean-field dynamics of the many-body system in terms of a 
Schr\"odinger equation for an effective single-particle wavefunction. Such a correspondence can be made precise in the Bethe Ansatz setting. The energy spectrum given by (\ref{nrg}) subject to the Bethe Ansatz equations (\ref{bae}) coincides with the quasi-exactly solvable sector of a one-dimensional Schr\"odinger equation with a particular potential.  Of the many solutions that (\ref{bae}) admits, it will be rigorously shown that the roots   
associated with the ground-state of the system with $A>0$, $\Omega>0$ lie on the positive real-axis of the complex plane. Moreover this is the unique solution with this property \cite{fn}.

Consider a general second-order ODE eigenvalue problem satisfied by an $m^{\text{th}}$-order polynomial $Q(u)$:
\begin{equation}
a(u)Q{''}(u)+b(u)Q'(u)+c(u)Q(u)=EQ(u)
\label{ODEform}
\end{equation}
First we write the polynomial $Q(u)$ with roots $\{v_p\}_{p=1}^m$ in the factorised form 
\begin{equation*}
Q(u)=\prod_{p=1}^{m}(u-v_{p}).
\end{equation*}
Evaluating (\ref{ODEform}) at the root $v_q$ leads to the Bethe Ansatz equations  
\begin{align}
\frac{b(v_q)}{a(v_q)}&=-\frac{Q''(v_q)}{Q'(v_q)} \nonumber \\
&= \sum_{p\neq q}^{m}\frac{2}{v_p-v_q},\hspace{1cm}q=1,2,...,m.
\label{BAeqns}
\end{align}
Hence, the roots of the polynomial must satisfy the system of coupled equations (\ref{BAeqns}) if $Q(u)$ is a solution to (\ref{ODEform}).

The  solutions of (\ref{ODEform}) with eigenvalue $E$ may be mapped to solutions of a Schr\"odinger equation 
\begin{equation}
\label{schro}
\frac{-{\rm d}^2\psi(x)}{{\rm d}x^2}+V(x)\psi(x)=E\psi(x)
\end{equation}
with the same eigenvalues,
by mapping the polynomial solution of (\ref{ODEform}) to a wavefunction of (\ref{schro}) via 
\begin{align}
\psi(x)=e^{f(x)}{Q(u(x))}.
\label{wavefn}
\end{align}
Substituting into the Schr\"odinger equation gives the following relations to be satisfied
\begin{subequations}
\label{mapp}
\begin{align}
a(u(x))&=-\left(\frac{{\rm d}u}{{\rm d}x}\right)^2, \label{a}\\
b(u(x))
&= -\frac{{\rm d}^2u}{{\rm d}x^2}-2\left(\frac{{\rm d}u}{{\rm d}x}\right)^2\frac{{\rm d}f}{{\rm d}u}, \label{b}\\
c(u(x))&=V(x)-\frac{{\rm d}^2f}{{\rm d}x^2}-\left(\frac{{\rm d}f}{{\rm d}x}\right)^2.\label{c}
\end{align}
\end{subequations}
The Bethe Ansatz solution of (\ref{ham}) derived in \cite{sl15} made use of the correspondence with the differential equation  
\begin{align}
&\left(Au^2 +4\Omega u \right) Q''(u) +\left(Bu+\Omega(4k+2-u^2) \right) Q'(u) \nonumber \\
&\qquad\qquad  +\left(C+\Omega Mu  \right)Q(u) =EQ(u), 
\label{ode}
\end{align}
for which it can be identified that 
\begin{align*}
a(u)&= A u^2 +4 \Omega u, \\
b(u)&= Bu+\Omega(4k+2-u^2),  \\
c(u)&= C+\Omega Mu .
\end{align*}
Then, from (\ref{a}), 
\begin{align*}
\frac{{\rm d}u}{{\rm d}x}&=\sqrt{-Au^2-4\Omega u}
\end{align*}
has the solution
\begin{align*}
u(x)&=\frac{2\Omega}{A}\left(\cosh(\sqrt{-A}x)-1\right). 
\end{align*}
Next use (\ref{b}) to evaluate 
\begin{align*}
\frac{{\rm d}f}{{\rm d}x}&=-\frac{1}{2}\left(b(u(x))+\frac{{\rm d}^2 u}{{\rm d}x^2}\right)\frac{{\rm d}x}{{\rm d}u} \\
&=\frac{K\sqrt{-A}}{2\sinh(\sqrt{-A}x)} 
\end{align*}
with 
\begin{align*}
K&=2k+{1}-\frac{B}{A} + \frac{(B-A)}{2A}\left(\cosh(\sqrt{-A}x)\right) \\
&\qquad -\frac{4\Omega^2}{A^2}\left(\cosh(\sqrt{-A}x)-1\right)^2.
\end{align*}
From here the potential is computed through (\ref{c})
which gives the explicit form
\begin{align}
V(x)=\frac{F(x)}{G(x)}
\label{pot}
\end{align}
where
\begin{align*}
F(x)&=  (12 \Omega^2 +8\Omega^2k-B^2)A^2(\cosh(\sqrt{-A}x)-1) \\
&\quad +A^4  (3-\cosh(\sqrt{-A}x)) \\
&\quad+4\Omega^4(1-\cosh(\sqrt{-A}x))^3 \\
&\quad +4AB\Omega^2(\cosh(\sqrt{-A}x)-1)^2 \\
&\quad +2A^3B(\cosh(\sqrt{-A}x)-2) \\
&\quad +4A^3C(\cosh(\sqrt{-A}x)+1) \\
&\quad +8A^2M\Omega^2(\cosh^2(\sqrt{-A}x)-1)\\
& \quad +8A^4  k-4A^3Bk, \\
G(x)&= 4A^3 (\cosh(\sqrt{-A}x)+1).
\end{align*}
Note that the potential is well-defined in the limit $A\rightarrow 0$, yielding the polynomial form
\begin{align*}
V(x)&=C - \frac{B(1+2k)}{4}  \\
&\qquad + \left( \frac{B^2 }{16} - \frac{(3+2k)\Omega^2}{4} - M\Omega^2\right) x^2  \\
&\qquad + \frac{B\Omega^2 x^4}{8} + \frac{\Omega^4 x^6}{16}.
\end{align*} 

 
\begin{figure}
\centering
{\includegraphics[width=9cm]{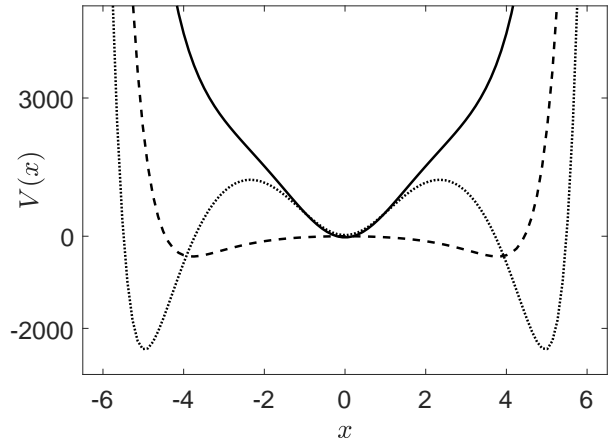}}\\
\caption{Illustrative examples of the potential (\ref{pot}). The parameter choices $k=0$, 
$M=20$, $A=-1$, $C=0$ and 
$\Omega=-1$ are fixed. The curves shown correspond to the choices  $B=100$ (solid line), $B=0$ (dash line), and $B=-100$ (dot line).}
\label{fig1}
\end{figure}
Now it will be shown that the ground state roots for $A>0$ and $\Omega>0$ lie on the positive real-axis, following a line of reasoning adopted from \cite{lz09}.   
First consider the case where $A<0$ and $\Omega<0$. It is seen that
$$\lim_{x\rightarrow \pm \infty}V(x)= \infty$$
and $V(x)$ is continuous and bounded below (see e.g. Fig. 1), in which case the oscillation theorem \cite{bs91} is applicable. Specifically, the wavefunction for the $m^{\rm th}$ energy level has $m$ real roots.  For a given value of $M$ and $k$, the first $M+1$ wavefunctions have the form (\ref{wavefn}) so the potential (\ref{pot}) is said to be {\it quasi-exactly solvable} \cite{uz92,u94}, with a one-to-one correspondence between this low energy spectrum and the full spectrum of (\ref{ham}) in the sector with $N=2M+k$. 

Consider next the highest energy state in the quasi-exactly solvable sector, which necessarily has $M$ real roots. 
The roots $v_j$ associated with each of the linear factors 
\begin{align*}
u-v_j=\frac{2\Omega}{A}\left(\cosh(\sqrt{-A}x)-1\right)-v_j
\end{align*}
in (\ref{wavefn}) must satisfy $v_j \geq 0$. These roots are also associated with the highest energy state of (\ref{ham}). This same state  is the ground state for $-H$, hence it is established that the roots of (\ref{bae}) for ground-state of (\ref{ham}) with $A>0$, $\Omega>0$ lie on the positive real-axis of the complex plane. That this is the unique solution with this property follows from the fact that the eigenspaces of (\ref{schro}) are one-dimensional. Hereafter, the analysis will be restricted to $A>0$, which will be referred to as the {\it repulsive} case.  


%
%

\section{Continuum limit and singular integral equation}
Eqs. (\ref{nrg},\ref{bae}) provide the Bethe Ansatz solution of (\ref{ham}) in {\it Richardson--Gaudin} form. This form facilitates the use of approximation by a singular integral equation in the continuum limit \cite{rsd02,admor02,bt07,lz09,lm15}, which is the approach that will be taken below. In the limit $M\rightarrow \infty$ a root density $\rho(v)$, with support on an interval $[\mathfrak{a},\mathfrak{b}]\subseteq (0,\infty)$, is introduced. The root density is required to be a solution of the continuum limit of (\ref{bae}), viz. the singular integral equation
\begin{align}
\lim_{M\rightarrow\infty}\frac{f(v)}{M}=P\int_\mathfrak{a}^\mathfrak{b} \frac{2\rho(w)}{w-v}\, {\rm d}w,
\label{sie}
\end{align}
where $P$ denotes the Cauchy principal value of the integral, subject to 
\begin{align}
\int_\mathfrak{a}^\mathfrak{b} \rho(w)\,{\rm d}w=1
\label{norm}
\end{align}
and 
\begin{align*}
f(v)&= \frac{Bv+\Omega(4k+2-v^2) }{Av^2 +4\Omega v }  \\
&=\frac{C_1}{v+4\Omega A^{-1}}+\frac{C_2}{v}-\frac{\Omega}{A},  
\end{align*}
where
\begin{align*}
C_1&=\frac{B}{A}+\frac{4\Omega^2}{A^2}-\frac{2k+1}{2},  \\
C_2&=\frac{2k+1}{2}.
\end{align*}  
Set $h=4\Omega A^{-1}$. Take the root density to be of the form 
\begin{align}
\rho(v)=\sqrt{(\mathfrak{b}-v)(v-\mathfrak{a})}\left(\frac{D}{v+h}+\frac{E}{v}  \right).
\label{dens}
\end{align}
Substituting (\ref{dens}) into (\ref{sie}) and  using the integral identities
\begin{align*}
P\int_\mathfrak{a}^\mathfrak{b} \frac{\sqrt{(\mathfrak{b}-w)(w-\mathfrak{a})}}{w(w-v)}\, {\rm d}w
&=\pi\left(\frac{\sqrt{\mathfrak{a}\mathfrak{b}}}{v}-1\right) \\
P\int_\mathfrak{a}^\mathfrak{b} \frac{\sqrt{(\mathfrak{b}-w)(w-\mathfrak{a})}}{(w+h)(w-v)}\, {\rm d}w
&=\pi\left(\frac{\sqrt{(\mathfrak{a}+h)(\mathfrak{b}+h)}}{v+h}-1\right)
\end{align*}
leads to the following identifications: 
\begin{align}
2\pi MD\sqrt{(\mathfrak{a}+h)(\mathfrak{b}+h)}&=C_1,  \label{e1}\\
2\pi ME \sqrt{\mathfrak{a}\mathfrak{b}}&=C_2,  \label{e2}\\
8\pi M (D+E)&=h. \label{e3}
\end{align}
Moreover, substituting (\ref{dens}) into (\ref{norm}) and using 
\begin{align}
\int_\mathfrak{a}^\mathfrak{b} \frac{\sqrt{(\mathfrak{b}-w)(w-\mathfrak{a})}}{w}\, {\rm d}w
&=\frac{\pi}{2}\left(\sqrt{\mathfrak{b}}-\sqrt{\mathfrak{a}}\right)^2, \label{int1} \\
\int_\mathfrak{a}^\mathfrak{b} \frac{\sqrt{(\mathfrak{b}-w)(w-\mathfrak{a})}}{w+h}\, {\rm d}w
&=\frac{\pi}{2}\left(\sqrt{\mathfrak{b}+h}-\sqrt{\mathfrak{a}+h}\right)^2 \label{int2}
\end{align}
yields
\begin{align}
1&=\frac{\pi D}{2}(\sqrt{\mathfrak{b}+h}-\sqrt{\mathfrak{a}+h})^2+\frac{\pi E}{2}(\sqrt{\mathfrak{b}} - \sqrt {\mathfrak{a}})^2.
\label{e4}
\end{align}
The constants $D$ and $E$ can be eliminated from the four equations (\ref{e1},\ref{e2},\ref{e3},\ref{e4}), leaving two equations. Introducing the notations
\begin{align*}
\alpha&= \frac{\sqrt{2N}}{\Omega}\left(\frac{U_{aa}}{2}-\frac{U_{bb}}{8}+\frac{\mu_a}{2N}-\frac{\mu_b}{4N}\right) ,\\
\lambda&= \frac{\sqrt{2N}}{\Omega}\left(\frac{U_{aa}}{2}-\frac{U_{ab}}{4}+\frac{U_{bb}}{8}  \right), \\
\kappa&=\sqrt{\mathfrak{a}\mathfrak{b}} , \\
\chi&= \sqrt{(1+\mathfrak{a}h^{-1})(1+\mathfrak{b}h^{-1})}, \\
Y
&=N+4\lambda (\alpha-\lambda) N+4\lambda^2, \\
Z&= 2\lambda (2k+1) \sqrt{2N} 
\end{align*}
these remaining two equations may be compactly expressed as
\begin{align}
N&=Y\chi^{-1}+Z\kappa^{-1}, \label{e5} \\
2N(4\lambda^2 +4\lambda\alpha +1)&= N\chi^2-2\lambda^2 \kappa^2 +Y\chi^{-1}-Z\kappa^{-1}. 
\label{e6}
\end{align}
Note that the condition $A>0,$ $\Omega>0$ means that $\lambda>0$.

The ground state energy, denoted $E_0$, can be expressed through the variables defined above. From (\ref{nrg}), we have the continuum limit approximation
\begin{align}
E_0&= U_{aa}N^2+\mu_a N -\Omega M\int_\mathfrak{a}^\mathfrak{b} v\rho(v)\,{\rm d}v \nonumber \\
&=U_{aa}N^2+\mu_a N + \frac{\Omega \kappa^2}{16\lambda }\left(\frac{N}{2}\right)^{1/2}  \nonumber \\
&\qquad \qquad +\frac{ \Omega Y  }{32\lambda^3}\left(\frac{N}{2}\right)^{1/2}\left(\chi-2+\chi^{-1}\left(1-\frac{\lambda^2\kappa^2}{2N}\right)\right) \nonumber \\
&\qquad \qquad -\frac{\Omega}{64\lambda^3}\left(\frac{N}{2}\right)^{3/2}\left(\chi^2-1-\frac{2\lambda^2 \kappa^2}{N}\right)^2 \label{connrg}
\end{align}
 which is obtained by using (\ref{e1},\ref{e2},\ref{e3},\ref{int1},\ref{int2},\ref{e4}) and 
\begin{align*}
\int_\mathfrak{a}^\mathfrak{b} \sqrt{(\mathfrak{b}-w)(w-\mathfrak{a})}\, {\rm d}w
&=\frac{\pi}{8}\left(\mathfrak{b}-\mathfrak{a}\right)^2.
\end{align*}

The parametrisations for  $\lambda$ and $\alpha$ were previously introduced in \cite{stfl06}, arising through an analysis of a classical analogue of the system. Through a study of the bifurcation of the classical fixed points, it was found that there was a transition line given by 
\begin{align}
\lambda=\alpha-1. \label{line}
\end{align}  
For $\lambda>0$ it was subsequently determined in \cite{sflmd10} that (\ref{line}) is a quantum phase transition boundary line. This conclusion was drawn from numerical studies of the energy gap, entanglement, and fidelity. The  objective below is derive certain correlation functions, through analysis of (\ref{e5}) and (\ref{e6}), which confirm this claim with complementary analytic results.

\section{Asymptotics for $N\rightarrow\infty$}
The leading order behavior of (\ref{e5},\ref{e6}), as $N\rightarrow \infty$,  may be exactly computed following the methods employed in \cite{lm15}. First consider the case where $\kappa$ has leading order term scaling as $\sqrt{N}$. Eqs. (\ref{e5},\ref{e6}) then yield
\begin{align}
\kappa^2&\sim 8N(\lambda-\alpha-1)(\lambda-\alpha+1),  \label{e7}\\
\chi&\sim 4\lambda(\alpha-\lambda)+1. \label{e8}
\end{align}
The requirement that $\kappa^2\geq 0$ subsequently imposes the restriction $\lambda\leq \alpha-1$. In this instance substituting (\ref{e7},\ref{e8}) into (\ref{connrg}) gives 
\begin{align}
E_0=\frac{U_{bb}N^2}{4}+\frac{\mu_b N}{2},
\label{connrgmol}
\end{align}
indicative of a pure molecular phase in the limit $N\rightarrow\infty$.

Alternatively, for the case $\lambda\geq \alpha-1$, consider  $\kappa$ has leading order term scaling as $1/\sqrt{N}$. Then 
\begin{align}
\kappa&= \frac{Z\chi}{N\chi-Y},  
\label{e9}
\end{align}
while to leading order $\chi$ is found to satisfy the following cubic equation
\begin{align}
\chi^3+p\chi+q=0 \label{cubic}
\end{align}
where
\begin{align*}
p&=-8\lambda^2-8\lambda \alpha-3  , \\
q&=-8\lambda^2+8\lambda\alpha +2  .
\end{align*}
Solutions of (\ref{cubic}) are given by 
\begin{align}
\chi=2\sqrt{\frac{-p}{3}}\cos\left(\frac{\theta+2l\pi}{3}\right), \qquad l=0,1,2
\label{roots}
\end{align}
where
\begin{align*}
\tan \theta&=\sqrt{\frac{-27q^2-4p^3}{27q^2}}. 
\end{align*}
Of the three solutions to (\ref{cubic}), the correct choice is that which reproduces (\ref{connrgmol}) in the limit $\lambda\rightarrow \alpha-1$. Setting $\lambda=\alpha-1$, which is equivalent to $p=-q^2/4-2$, (\ref{cubic}) factorises as
\begin{align*}
\left(\chi-\frac{q}{2} \right) &\left(\chi+\frac{q}{4}-\frac{1}{4}\sqrt{q^2+32} \right) \\
\times &\left(\chi+\frac{q}{4}+\frac{1}{4}\sqrt{q^2+32} \right)=0.
\end{align*}
From this limiting case it is then deduced that $l=0$, giving $\chi=q/2$, is the appropriate choice to be made in (\ref{roots}) with the arctangent chosen such that 
\begin{align*}
\cos(\tan^{-1}(x))=-\frac{1}{\sqrt{1+x^2}}. 
\end{align*} 

Additionally, it needs to be confirmed that (\ref{cubic}) is only applicable when $\lambda\geq \alpha-1$. To verfiy this, first observe that the denominator term of (\ref{e9}) imposes 
\begin{align*}
\chi\geq \frac{Y}{N} \sim \frac{q}{2}.
\end{align*}
Construct the function 
\begin{align*}
f(\alpha,\lambda)&=2\chi-q, 
\end{align*} 
such that $f(\lambda+1,\lambda)=0$ as $N\rightarrow \infty$. The solution for the root density is only valid while 
$f(\alpha,\lambda)\geq 0$. Now from (\ref{cubic})
\begin{align*}
\frac{\partial \chi}{\partial \alpha}&=8\frac{\lambda\chi-\lambda}{3\chi^2+p}
\end{align*}
so 
\begin{align*}
\frac{\partial f}{\partial \alpha}(\alpha,\lambda)&=16\frac{\lambda\chi-\lambda}{3\chi^2+p} -8\lambda ,\\
\frac{\partial f}{\partial \alpha}(\lambda+1,\lambda)&=-\frac{32\lambda^2(4\lambda+1)}{(4\lambda+1)^2-1}.
\end{align*}
Bearing in mind that $\lambda >0$ it is seen from 
$$\frac{\partial f}{\partial \alpha}(\lambda+1,\lambda)<0,$$ 
along with $f(\lambda+1.\lambda)=0$, that the requirement $f(\alpha,\lambda) \geq 0$ imposes the constraint $\lambda \geq \alpha - 1$. Finally, for this phase the energy expression from (\ref{connrg}) gives 
\begin{align*}
E_0&= U_{aa}N^2 +\mu_a N +\Omega\left(\frac{N}{2}\right)^{3/2} \Theta 
\end{align*}
where 
\begin{align*}
\Theta=\frac{1}{64\lambda^3}(p\chi^2+3q\chi-2p-4q-1).
\end{align*} 

\section{Correlation functions}

Having obtained an explicit expression for the ground-state energy, several correlation functions can be computed through use of the Hellmann-Feynman theorem. The first of these is the expectation value of the atomic fraction, which is given by
\begin{align*}
\frac{\langle N_a \rangle}{N}&=\frac{1}{N}\frac{\partial E_0}{\partial \mu_a}.
\end{align*}
For $\lambda\leq \alpha -1$ 
\begin{align*}
\frac{\langle N_a \rangle}{N}=0,
\end{align*}
while for $\lambda \geq \alpha-1$
\begin{align*}
\frac{\langle N_a \rangle}{N}
&=1+\Omega\left(\frac{N}{2}\right)^{1/2}\frac{\partial \Theta}{\partial \alpha}\frac{\partial \alpha}{\partial \mu_a} \nonumber \\
&=1  -2^{-5}K
\end{align*}
where
\begin{align}
K=\frac{(6-4p)\chi^2+(8p-6q)\chi+12q+2p}{\lambda^2(3\chi^2+p)}. 
\label{K}
\end{align}
A graphical representation of this quantity is given in Fig. 2.
\begin{figure}[t]
\centering
{\includegraphics[width=9cm]{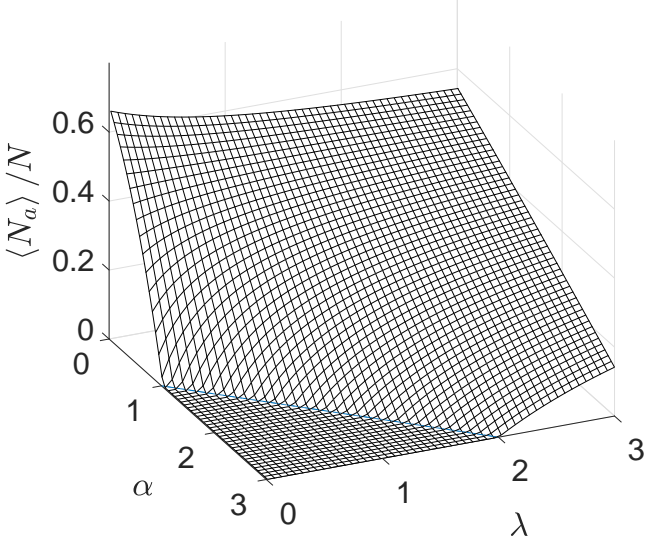}}\\
\caption{Surface plot of the ground-state atomic fraction $\left<N_a\right>/N$ as a function of $\lambda$ and $\alpha$. The ground-state atomic fraction is zero in the region $\lambda\leq \alpha-1$, which is the molecular phase. It takes non-zero values in the mixed phase $\lambda\geq \alpha-1$.}
\label{fig2}
\end{figure}

Moreover, several other correlation functions are also computable using the same approach. Specifcally, for $\lambda\leq \alpha -1$ it is found that 
\begin{align*}
\frac{\langle N_b \rangle}{N}&=\frac{1}{2}, \\
\frac{\langle N_a^2\rangle}{N^2}&=0, \\
\frac{\langle N_b^2 \rangle}{N^2}&=\frac{1}{4}, \\
\frac{\langle N_a N_b \rangle}{N^2}&= 0,\\
\frac{\langle a^\dagger a^\dagger b + a a b^\dagger \rangle}{N^{3/2}}&= 0
\end{align*}
while for $\lambda\geq \alpha -1$
\begin{align*}
\frac{\langle N_b \rangle}{N}&=2^{-6}K,\\
\frac{\langle N_a^2\rangle}{N^2}&=1-2^{-9}J-2^{-5}K, \\
\frac{\langle N_b^2 \rangle}{N^2}&= 2^{-7}K-2^{-11}J, \\
\frac{\langle N_a N_b \rangle}{N^2}&=  2^{-10}J, \\
\frac{\langle a^\dagger a^\dagger b + a a b^\dagger \rangle}{N^{3/2}} 
&=2^{-3/2}(\Theta+2^{-7}\lambda J+2^{-3}\alpha K )
\end{align*}
with 
\begin{align*}
J
&=\lambda^{-4}((3p-q-7)\chi^2 
+3(3+3q-p)\chi -2p-10q-4) \\
&\qquad +\lambda^{-4}(2p\chi+3q)\left(\frac{(3p+q+7)\chi +p+3q-3}{3\chi^2+p}\right)
\end{align*}
and $K$ given by (\ref{K}).
Following \cite{zlm03} the coherence correlator $C$ is defined as
\begin{align}
C=-\frac{1}{2}\frac{\langle a^\dagger a^\dagger b + a a b^\dagger \rangle}{N^{3/2}}.
\label{cf}
\end{align}
A graphical representation of this quantity is given in Fig. 3.

Finally, it can be verified through the above formulae that the variances are all zero, that is 
\begin{align*}
\frac{\langle N_a^2 \rangle}{N^2}&=  \frac{\langle N_a \rangle^2}{N^2}, \\
\frac{\langle N_b^2 \rangle}{N^2}&=  \frac{\langle N_b \rangle^2}{N^2},\\
\frac{\langle N_a N_b \rangle}{N^2}&=  \frac{\langle N_a \rangle \langle N_b \rangle}{N^2}. 
\end{align*}
\begin{figure}[t]
\centering
{\includegraphics[width=9cm]{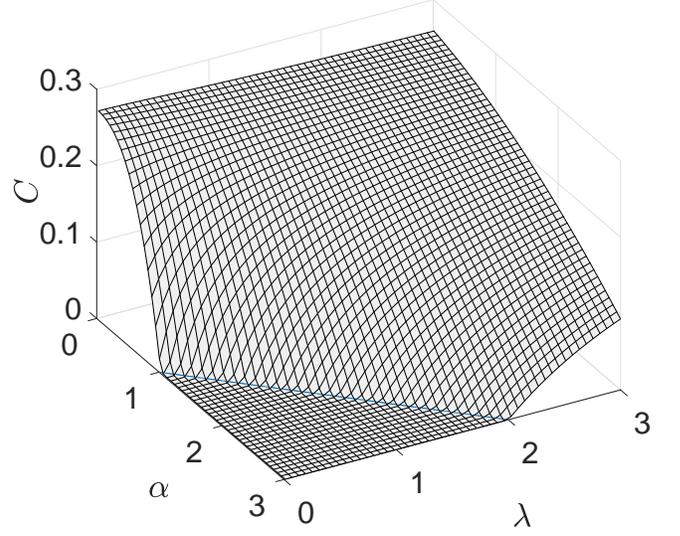}}\\
\caption{Surface plot of the ground-state coherence correlator $C$ as a function of $\lambda$ and $\alpha$. The ground-state coherence correlator is zero in the region $\lambda\leq \alpha-1$, which is the molecular phase. It takes non-zero values in the mixed phase $\lambda\geq \alpha-1$.}
\label{fig3}
\end{figure}

\section{Conclusion}

In this study we have re-examined the boundary between the molecular and mixed phases for the Hamiltonian (\ref{ham}) in the repulsive case $\lambda \geq 0$. 
The quantum phase boundary line $\lambda=\alpha-1$ was first identified in \cite{stfl06} via  a bifurcation analysis of fixed points for a corresponding classical system. Numerical studies undertaken in \cite{sflmd10}, which studied quantities such as entanglement and fidelity, provided further supporting evidence. Here we have obtained an exact, analytic result for the ground-state energy in the limit as $N\rightarrow\infty$ which confirms the quantum phase boundary. Through this result several ground-state correlation functions can also be evaluated. Illustrative examples were provided through the expectation value of the ground-state atomic fraction, as depicted in Fig. 2, and the fluctuation of that quantity.     
 
For future work it remains to extend the analysis to the entire $(\alpha,\lambda)$ plane of the coupling space. The bifurcation analysis of \cite{stfl06} points towards the existence of additional ground-state phases, but the characterisation of them is not yet clear. For the attractive case $\lambda<0$ there appears to be no purely atomic phase, but two distinct mixed phases. It is anticipated that further development of exact, analytic results will enable new insights to be gained, leading  to a better understanding of the system. 
 
\section*{Acknowledgments}
\noindent
This work was supported by the Australian Research Council through Discovery Project DP150101294.





\begin{thebibliography}{99}
\bibitem{fn} The unitary transformation $b \mapsto -b$ is equivalent to a change of sign for $\Omega$. Note that (\ref{nrg}) and (\ref{bae}) are invariant when this is also accompanied by the change $v_j \mapsto -v_j$. So in the case that $A>0$ and $\Omega<0$ the roots   
associated with the ground-state of the system lie on the negative real-axis, and is the unique solution with this property.
\end{thebibliography}
\end{document}